# The Microlensing and Halo Models of the Galaxy

Yukitoshi Kan-ya, Ryoichi Nishi,
*Department of Physics, Kyoto University*
*Kyoto 606-01, Japan*
and
Takashi Nakamura
*Yukawa Institute for Theoretical Physics, Kyoto University*
*Kyoto 606-01, Japan*

## Abstract

We investigated the dependence of the optical depth $\tau$ of the microlensing events on model parameters of the Galactic halo. We only consider Galactic mass models in which the rotation curve inside the Sun is compatible with the observation and LMC is bound to the Galaxy. It is found that $\tau$ varies up to a factor 2.5 from the standard spherical and flat rotation halo model. This implies that only the most centrally concentrated halo model can be consistent with the observation if the halo consists of only MACHOs. We also calculate the power $x$ of IMF of MACHO consistent with Tyson's CCD survey as well as Bahcall *et al.* 's observation by *HST*. It is found that $x$ is greater than 5.

## 1 Introduction

Gravitational microlensing events are detected recently by three collaborations: MACHO (Alcock *et al.* 1993 [3]) and EROS (Aubourg *et al.* 1993 [8], Aubourg *et al.* 1995 [9]) for LMC events, as well as OGLE (Udalski *et al.* (1993) [42]) and MACHO (Bennett *et al.* 1994 [13]) for bulge events. From these results we can discuss about the nature of the missing mass in our Galaxy (Paczyński 1986 [30]).

The key quantity in this problem is the optical depth $\tau$ of the microlensing. This quantity is the instantaneous probability that the event is occurring when we observe background stars randomly. Observationally $\tau$ is derived from the number of the observed stars, the mean event duration and the event rate (Paczyński 1986[30]). For LMC events,

MACHO collaboration concluded that $\tau = 8.0^{+14}_{-6} \times 10^{-8}$ in 68% confidence level(Alcock et al. (1995) [5]). On the other hand, several authors (e.g. Paczyński 1986 [30] Griest 1991 [24]) estimated $\tau$ for the halo model with flat rotation curve in the outer region of the Galaxy and obtained $\tau \simeq 5 \times 10^{-7}$, which suggests a MACHO fraction $f = 0.2^{+0.33}_{-0.14}$ and all of the missing mass of our Galaxy may not be MACHOs. For bulge events, the observational value of $\tau$ is $\simeq 3 \times 10^{-6} \epsilon^{-1}$ (Udalski 1994 [43]) by OGLE collaboration, where $\epsilon$ is the efficiency of the observation, and $\tau = 3.0^{+1.5}_{-0.9} \times 10^{-6}$ by MACHO collaboration (Bennett 1994 [13]). Paczyński (1991) [31], Griest et al. (1991) [25] and Kiraga & Paczyński (1994) [26] estimated $\tau$ for the bulge as $0.1 \sim 1 \times 10^{-6}$, which is at least factor 3 smaller than the observational data, that is, $f \sim 3$. The fraction, $f$, however, depends on the theoretical estimate of $\tau$ i.e. models of our Galaxy. So it is important to estimate the model dependence of $\tau$. The main purpose of this paper is to discuss theoretical $\tau$ in more detail.

Although the model with the flat rotation curve is frequently taken as the mass distribution in the halo, at present we can only say that the Galactic rotation curve is essentially flat *only inside the solar neighborhood* (e.g. figure 2 in Fich & Tremaine 1991 [18]). In the outer region we have no definite rotation curve at present although we can impose some constraints as will be discussed in section 2. Many other spiral galaxies have flat rotation curve up to the outer most region. However, it is reported that spiral galaxies with its exponential disk scale length less than 3.5kpc have declining rotation curves (Casertano & van Gorkom 1991 [15]). For our Galaxy its scale length seems to be marginal, i.e., $\sim$ 3.5kpc. Our Galaxy may have non-flat rotation curve beyond solar neighborhood.

The shape of the halo is another point to be considered. It is suggested from N-body simulations of the galaxy formation that the halo may be nonspherical (e.g. Aarseth & Binney 1978, [1], Aguilar & Merritt 1990 [2], Binney 1994 [14] and references therein). Sackett & Gould (1993) [36] and Frieman & Scoccimarro (1994) [19] discussed that the ratio of the optical depth toward SMC to LMC is a good probe for the shape of the halo. In this paper, we also investigate the dependence of $\tau$ on the shape.

As a model of the halo we take the model of Evans (1994) [17] that is a power-law model with a rising, flat or falling rotation curve. We impose some constraints in this model and calculate the dependence of the optical depth $\tau$ on model parameters. The same model has been used by Alcock et al. (1994) [4]. They concluded that $\tau$ changes up to a factor 10. In this paper we use more stringent constraints than theirs, which will be shown later.

MACHOs may be the low-mass stars. Richer & Fahlman (1992) [33] suggested that the IMF of the low-mass star less massive than $\sim 0.5 M_\odot$ in the Galactic spheroid stars is much steeper than the Salpeter's IMF. Bahcall et al. (1994) [12] observed recently halo stars in a high-latitude region by *HST* and concluded that if the dark halo consists of low-mass stars they must have mass less than hydrogen-burning limit. Using data of the CCD survey of Tyson (1988) [40] and *HST* observation by Bahcall et al. (1994), we will discuss on the constraints to the power of IMF assuming the power law IMF.

The plan of this paper is as follows. In section 2 we will show the Galactic model and impose the constraint on it. In section 3 we will derive the dependence of $\tau$ on our model parameter. In section 4 we discuss the power of IMF consistent with the observation.

Section 5 is devoted to discussions.

## 2  The Galactic model

As a spherical or spheroidal Galactic model we take an axisymmetric power-law model (Evans 1994 [17]). In this model the gravitational potential is given in the cylindrical coordinate $(R, \phi, z)$ as

$$\Psi = \frac{v_0^2 R_c^\beta / \beta}{(R_c^2 + R^2 + z^2 q^{-2})^{\beta/2}}, \qquad (1)$$

where $v_0$, $R_c$, and $q$ are the normalization of potential, the core radius and the axis ratio of equipotential, respectively. The rotation velocity in the equatorial plane is

$$v_c = \left( \frac{v_0^2 R_c^\beta R^2}{(R^2 + R_c^2)^{(\beta+2)/2}} \right)^{\frac{1}{2}} \sim R^{-\frac{\beta}{2}} \text{ as } R \to \infty. \qquad (2)$$

The mass density of the halo is given as

$$\rho_h = \frac{v_0^2 R_c^\beta}{4\pi G q^2} \frac{R_c^2(1 + 2q^2) + R^2(1 - \beta q^2) + z^2(2 - (1+\beta)q^{-2})}{(R_c^2 + R^2 + z^2 q^{-2})^{(\beta+4)/2}}. \qquad (3)$$

The asymptotic ellipticity $e$ of the edge-on isophotal contours for $R \to \infty$ is given as

$$e = 1 - \left[ \frac{q^{3+\beta}(1 - \beta q^{-2})}{1 - \beta q^2} \right]^{\frac{1}{1+\beta}}. \qquad (4)$$

For the other components of the Galaxy we take the model by Bahcall, Schmidt, & Soneira (1982) [11] with the exponential disk, the $r^{\frac{1}{4}}$ spheroid and the central mass concentration. The surface density of known matter in the Galactic disk at the solar neighborhood is $\Sigma_{id} = 48 \pm 8 M_\odot \text{pc}^{-2}$ ([7]). Bahcall, Flynn & Gould (1992) [10] have claimed that the total surface density $\Sigma_0 = 88 M_\odot \text{pc}^{-2}$, implying that there is large amount of the disk DM. On the other hand, Kuijken & Gilmore (1989) [28] concluded that $\Sigma_0 = 46 \pm 9 M_\odot \text{pc}^{-2}$, and there is no evidence for the disk DM. Since the existence of disk DM is not established, we regard $\Sigma_0$, or the local mass density of the disk $\rho_{d0} = \Sigma_0 / 2z_h$, where $z_h (= 300 \text{pc})$ is the scale height of the disk, as a parameter. We consider two cases, the heavy disk (with disk DM) and the light disk (without disk DM), respectively. The density profile of the disk, the spheroid and the central mass concentration are respectively

$$\rho_d = \rho_{d0} \exp\left( -\frac{R - R_0}{R_d} - \frac{z}{z_h} \right) \qquad (5)$$

$$\rho_{sph} = 9 \times 10^{-5} M_\odot / \text{pc}^{-3} \left( \frac{r}{R_0} \right)^{-\frac{7}{8}} \exp\left[ -10.093 \left( \frac{r}{R_0} \right)^{\frac{1}{4}} + 10.093 \right] \qquad (6)$$

$$\times \left[ 1 - 0.08669 \Big/ \left( \frac{r}{R_0} \right)^{\frac{1}{4}} \right] \qquad (7)$$

$$\rho_{cen} = 7.6 \times 10^5 M_\odot / \text{pc}^{-3} \left[ \left( \frac{R}{1\text{kpc}} \right)^2 + 6.25 \left( \frac{z}{1\text{kpc}} \right)^2 \right]^{-0.9} \exp\left[ -\left( \frac{R}{1\text{kpc}} \right)^3 \right] \qquad (8)$$

where $r$, $R_0(=8.5\text{kpc})$, and $R_d(=3.5\text{kpc})$ are the distance from the Galactic center, the galactocentric radius of the sun, and the disk scale length, respectively. As the total mass density $\rho$ we take $\rho = \rho_h + \rho_d + \rho_{sph} + \rho_{cen}$. The rotation curve is determined by the sum of the gravitational potential of the above four components.

Now we have five parameters in our Galactic model: $R_c$ (the core radius of the halo), $v_0$ (the normalization of potential due to halo), $e$ (the asymptotic ellipticity of the halo), $\beta$ (the slope of the rotation curve), and $\rho_{d0}$ (the local mass density of the disk). We have, however, some observational constraints to these parameters:

1. Although as discussed in section 1 we have no definite rotation curve for $R > R_0$, for $R < R_0$ the nearly flat rotation curve is observed. We take this fact into account. We fix $v_c(R = R_0) = 225\text{km/s}$ and $v_c(R = 2\text{kpc}) = 200\text{km/s}$ (e.g. [18]) on the Galactic plane. This constraint determines the value of $v_0$ and $R_c$ for given values of $\beta$ and $\rho_{d0}$.

2. Kroupa, Röser & Bastian (1994) [27] have observed the proper motion of LMC and found that the mean galactocentric space motion vector is $(-279 \pm 165, -218 \pm 79, +85 \pm 122)\text{km/s}$. Adopting this proper motion, we require that LMC is gravitationally bound to the Galaxy. This restricts allowed range of $\beta$, $e$. Too large $\beta$ is not possible.

3. Thomas (1989) [39] observed the motion of the Galactic globular clusters in the halo and got the velocity dispersion along the line-of sight direction as $140 \pm 37\text{km/s}$ and the tangential direction as $223 \pm 132\text{km/s}$ for the globular clusters with galactocentric radius between $\sim 7\text{kpc}$ and $\sim 20\text{kpc}$. Our rotation curve must be consistent with this result.

Under the constraint 1 the rotation curve on the equatorial plane is shown in Figure 1. Note that this curve is independent of $q$ since the potential at z=0 is independent of $q$ (see equation 2). Using constraints 2 and 3, we have the allowed range of $\beta$ as approximately $-1 \leq \beta \leq 0.6$ for the heavy disk and $-0.8 \leq \beta \leq 0.6$ for the light disk. As for $e$, we consider E0~E7 halo ($0 \leq e \leq 0.7$) (Rix 1994 [34]). In reality, the constraint 3 does not play an important role.

The mass of the Galaxy in the above range of parameters is $2 \times 10^{11} M_\odot$ to $2 \times 10^{12} M_\odot$ within $r = 55\text{kpc}$. To $r = 100\text{kpc}$ it is $8 \times 10^{11} M_\odot$ to $8 \times 10^{12} M_\odot$. This value is consistent with the observational estimates (e.g. Ashman 1992 [7]).

## 3  The optical depth of microlensing

In this section we calculate the optical depth of microlensing in our model for LMC events and bulge events, respectively. The optical depth of the microlensing $\tau$ is the probability for MACHOs to be in the microlensing tube ([24]) and written as ([24], [30])

$$\tau = \int_0^{D_S} \frac{\rho(\mathbf{r}(D_L))}{m} \pi u_T^2 R_E^2(D_L) dD_L, \qquad (9)$$

Figure 1: Rotation curves on the equatorial plane of our Galactic model for specified values of $\beta$. (a) is for the heavy disk, while (b) is for the light disk.

where $D_L$ is the distance to the lens object from the Sun, $m$ is the mass of the object, $D_S$ is the distance to the source and $R_E = [(4Gm/c^2)(1 - D_L/D_S)]^{1/2}$ is the Einstein radius. We will calculate $\tau$ for $u_T = 1$.

In Figures 2 and 3 we show the cumulative contribution $f_c$ to $\tau$ defined by

$$f_c(D_L) = \int_0^{D_L} \frac{\rho(\mathbf{r}(D_L))}{m} \pi u_T^2 R_E^2(D_L) dD_L \Big/ \int_0^{D_S} \frac{\rho(\mathbf{r}(D_L))}{m} \pi u_T^2 R_E^2(D_L) dD_L \ . \qquad (10)$$

In Figure 2 we show $f_c$ for E0 halo. The contribution to $\tau$ mainly comes from the halo component. We also see that the mass distribution in the relatively neighbor region i.e. $D_L < 20$kpc from the sun mainly determines the optical depth. In Figure 3, we show $f_c$ for E6 halo. The increase of $e$ makes the contributing region more nearer and narrower. We show in Figure 4 $f_c$ for bulge events with the same notation as for LMC events. For bulge events, the contribution is almost independent of $\beta$ and $e$. The main contribution comes from the disk component for the heavy disk, while for the light disk the contribution from the disk and from the halo is comparable.

In Figure 5 we show $\tau$ to the direction toward LMC with $D_S = 55$kpc. From this figure we see that $\tau$ does not change more than a factor 2.5 from the model with the spherical halo and flat rotation curve ($e = 0, \beta = 0$), as we change $\beta$, $e$, and $\rho_{d0}$. This result comes from the fact that the rotation curve for $R < R_0$ almost fixes the structure of the halo within $D_L \leq 20$kpc (or $r \leq 20$kpc) where the contribution to $\tau$ is more than 60 %. We also see that $\tau$ does not depend on $e$ strongly. This is consistent with the results of Sackett & Gould (1993) [36] and Frieman & Scoccimarro (1994) [19], although they used a simpler model of the Galaxy than ours.

Using the Galactic model with the halo and disk models as ours, without the spheroid and the central component, Alcock *et al.* (1994) [4] concluded that $\tau$ changes of an order of magnitude. They imposed the constraint that the rotation velocity lies between 180km/s and 250km/s at $R_0$ and $2R_0$. But in their model it is not clear that the inner rotation curve is reproduced. Under present constraints in this paper, the rotation curve of the Galaxy for $r \leq 15$kpc is almost independent of parameters. Since this region has the dominant contribution to $\tau$ as we can see in Figure 2 and 3, the parameter dependence of $\tau$ is much smaller than that of Alcock *et al.* (1994)

We also show $\tau$ to the direction towards Baade's window with $D_S = R_0$ in Figure 6. $\tau$ is almost independent of $\beta$ and $e$ due to the constraints to the rotation curve. $\tau \simeq 1.2 \times 10^{-6}$ for the heavy disk and $\tau \simeq 8 \times 10^{-7}$ for the light disk. As mentioned in section 1, MACHO collaboration observed that the optical depth for bulge events is $3.0^{+1.5}_{-0.9} \times 10^{-6}$, which is a factor $3 \sim 6$ greater than the previous theoretical estimates of Paczyński (1991) [31] and Griest *et al.* (1991) [25] taking the stellar component of the disk. Our result implies that even if we take into account the halo and spheroid component, the optical depth is at least a factor 2.5 less than that observed.

Finally, we show the model dependence of the ratio of the optical depth for SMC and LMC, $\tau_{SMC}/\tau_{LMC}$, in Figure 7. $\tau_{SMC}/\tau_{LMC}$ increases as $\beta$ increases from 0. This is because of the following reason. The line of sight toward SMC passes through nearer to the Galactic center. For $\beta > 0$ the outer halo is not important as in the case of $\beta = 0$. On the other hand when $\beta < 0$ and $e > 0$ the compression of the less centrally concentrated halo along the z-direction makes the density in the dominant region for the microlensing increase specially for SMC, so that $\tau_{SMC}/\tau_{LMC}$ increases in this case also.

Figure 2: The cumulative contribution to the optical depth as a function of $D_L$ for LMC events for asymptotically E0 halo. The disk model is (a)heavy disk (b)light disk

Figure 3: The cumulative contribution to the optical depth as a function of $D_L$ for LMC events for asymptotically E7 halo. The disk model is (a)heavy disk (b)light disk

Figure 4: The cumulative contribution to the optical depth as a function of $D_L$ for bulge events for asymptotically E0 halo. The disk model is (a)heavy disk (b)light disk

Figure 5: The optical depth towards the direction of LMC as a function of $\beta$, (a) heavy disk and (b) light disk.

Figure 6: The optical depth of the microlensing towards the direction of Baade's window, as a function of $\beta$. (a) heavy disk and (b) light disk.

From Figure 7 we see for $e \leq 0.3$, $\tau_{SMC}/\tau_{LMC}$ does not depend on $\beta$ so that it may be a tracer of halo flattening, but for $e \geq 0.3$ the ratio changes more as a function of $\beta$. So from $\tau_{SMC}/\tau_{LMC}$, we obtain a certain combination of flatness and $\beta$. Sackett & Gould (1993) [36] and Frieman & Scoccimarro (1994) [19] used only the model with the flat rotation curve and concluded that $\tau_{SMC}/\tau_{LMC}$ is a good probe of the flattening of the halo if the halo-disk tilt angle is $\simeq 0$. Our result shows that even if the tilt angle is 0, we may not determine the flattening directly without the knowledge of $\beta$.

## 4  The stellar number count and the halo IMF

Now we calculate stellar number counts in our model and argue on the IMF in the halo. Richer & Fahlman (1992) [33] observed low mass stars down to $0.14 M_\odot$ and suggested that the slope of the IMF is steeper than Salpeter's one. More recently Bahcall *et al.* (1994) performed star count of high latitude field by *HST* and found fewer faint red stars. This suggests that the low-mass stars less massive than the hydrogen-burning limit, $0.08 M_\odot$ may dominates in the halo, if the dark halo or the part of the dark halo consists of the low-mass stars. This is consistent with the statistical mass estimate of MACHOs (Sutherland *et al.* 1994 [38]).

We now assume $\simeq 20\%$ of halo consists of MACHO (Alcock *et al.* 1995 [5]) and ask what its IMF would be. We assume power law IMF for MACHO. We use the data of Tyson's CCD survey(Tyson 1988 [40], Tyson & Seitzer 1988 [41]). He observed 12 high-latitude fields where there are no stars or galaxies brighter than $B_J = 20$ and $R = 19$ exist, and found about 50 stellar-like objects in a field with $2.6 \times 4.6 \mathrm{arcmin}^2$ wide. This number count provide a constraint to the halo IMF if the dark halo consists of stars and brown dwarfs.

If the fraction $f$ of the halo consists of MACHO in our model of the Galaxy the expected star counts $N$ in Tyson's CCD survey is calculated as

$$N = d\Omega \int_{D \geq D_{min}} D^2 dD \int dM \rho \phi \times f \qquad (11)$$

$$= d\Omega \int_{D \geq D_{min}} D^2 dD \int_{m \leq m_{lim}} dm \rho \phi \left(\frac{dL(M)}{dM}\right)^{-1} \frac{\partial L(m,D)}{\partial m} \times f, \qquad (12)$$

where $m$ is the apparent magnitude of the star, $D$ is the distance to the star, $d\Omega$ is the solid angle of the field, $D_{min} = 100$pc is the minimum distance from which we begin to count stars, $m_{lim}$ is the limiting magnitude, and $L$ is the V band luminosity of the star. IMF $\phi$ is normalized as $\int \phi dM = 1$. $L(M)$ is the mass-luminosity relation. This relation is derived by D'Antona (1987) [16] for the population II stars of $M \lesssim 1 M_\odot$ and approximately given as $L(M) \propto M^{2.7}$. Now we neglect the contribution of massive stars to $N$ and we assume $M < 0.08 M_\odot$ does not shine.

Figure 7: $\tau_{SMC}/\tau_{LMC}$ as a function of $\beta$ for various $e$. (a) is for the heavy disk and (b) is for the light disk.

Figure 8: The total number count $N$ as function of $\beta$ for various $x$ for the direction of Tyson's SGP ($l = 312.5°, b = -89.3°$) data assuming the heavy disk.

We use a power law IMF such as $\phi \propto M^{-(1+x)}$. For the disk stars we use the Scalo's IMF (Scalo 1986 [37]), which is flat, i.e., $x = -1$ for $0.08 M_\odot \leq M \leq 1 M_\odot$ and $x = 2.3$ for $1 M_\odot \leq M \leq 50 M_\odot$. For spheroid stars we take Richer & Fahlman's value $x = 3.5$ for $M_{min} \leq M \leq 1 M_\odot$. For dark halo, $x$ is a parameter and mass varies as $M_{min} \leq M \leq 1 M_\odot$. We put the low-mass end of IMF as $M_{min} = 0.01 M_\odot$. This value is suggested from Sutherland (1994)'s estimate of the mass of MACHOs and a star formation theory of Palla, Salpeter & Stahlar (1983) [32]. The observations of MACHO (Alcock *et al.* 1995 [5]) and EROS collaboration (Aubourg *et al.* 1995 [9]) suggest that the mass of MACHOs is not so much less than $0.1 M_\odot$.

In Figure 8 we show $N$ as a function of $\beta$ for given $x$ for the direction of Tyson(1988)'s SGP ($l = 312.5°, b = -89.3°$) data. $N$ is not sensitive to $\beta$ but strongly depends on $x$. This is because $N$ is determined mainly by the number of hydrogen burning limit stars in the halo at the distance of $D = 3 \sim 5$kpc. In this distance, the structure of the halo changes little as $\beta$ changes, as discussed in section 3. The number of the shining low-mass stars is very sensitive to $x$. For other Tyson(1988)'s fields the figure does not changes so much. Since Tyson (1988) observed about 50 stars in a field, we conclude that IMF's power in the halo is steep at least as $x \simeq 5$.

Recently, using *HST*, Bahcall *et al.* (1994) [12] observed a field of 4.4arcmin$^2$ wide, with $I < 25.3$ and $V < 25.6 - 0.3(V - I)$. They found five stars with $2 < V - I$ ($M \leq 0.4 M_\odot$ if brown dwarfs). Size of their field is about half of Tyson's one. Similar calculations show $x \geq 6$.

## 5  Conclusion

We investigated how Galactic mass models and the flattening of the halo affects $\tau$ under the assumption that the halo consists of only MACHOs. We found that the constraints of the inner rotation curve almost uniquely determines the structure of the region of the halo ($\leq 20$kpc) where the microlensing events mainly occur. As a result $\tau$ for LMC events varies at most a factor 2.5 from the standard spherical flat rotation curve model of $\tau \sim 4 \times 10^{-7}$. For $f \sim 1$, i.e., the halo consists of only MACHOs, the observation of MACHO collaboration $8^{+14}_{-6} \times 10^{-8}$ is consistent only if we take the most centrally concentrated model with $\beta \sim 0.6$, which means the rotation curve at $r \sim 20$kpc is declining.

We also showed that $\beta$-dependence of $\tau_{SMC}/\tau_{LMC}$ is large for $\epsilon \geq 0.3$. This implies that $\tau_{SMC}/\tau_{LMC}$ does not necessarily determine the flattening of the halo unless the rotation curve at $r \sim 20$kpc is determined definitely.

For the bulge events, $\tau$ cannot reach the observational value even in our models in which effects of the ellipticity and non flat rotation curve are included. This means that we must consider other effects such as a bar in the bulge. As Gould(1994a) [20] suggested, all of the matter inside the Sun may be in the thin disk and the LMC event may be by the stars in LMC. Anyway we need further observation and statistics (e.g. Gould (1994b) [21], Miyamoto & Yoshii (1994) [29], Gould (1994c) [22], Ansari *et al.* (1995) [6], Gould (1994d) [23]).

For LMC events, we calculated the power $x$ of IMF of MACHO consistent with Tyson's CCD survey as well as Bahcall *et al.* 's observation by *HST*. It is found that $x$ is greater than 5. This suggests that the halo IMF is essentially a $\delta$-function if the halo consists of only low mass stars. High value of $x$ is required even if $f \sim 0.2$ unless the microlens events are occurred near stars in LMC.

We thank Professor H. Sato for useful discussions. This Work was supported by Grant-in-Aid of Scientific Research of the Ministry of Education.

## References

[1] Aarseth, S. J., Binney, J. J. (1978) MNRAS., **185**,227

[2] Aguilar, L. A., Merritt, D. R. (1990) ApJ., **265**, 33

[3] Alcock, C. *et al.* (1993) *Nature*, **365**, 621

[4] Alcock, C. *et al.* (1994) astro-ph/9411019

[5] Alcock, C. *et al.* (1995) submitted to Phys. Rev. Lett., astro-ph/9501091

[6] Ansari *et al.* (1995) astro-ph/9502102

[7] Ashman, K. M. (1992) PASP., **104**, 1109

[8] Aubourg, E. *et al.* (1993) *Nature*, **365**, 623

[9] Aubourg, E. *et al.* (1995) astro-ph/9503021

[10] Bahcall, J. N., Flynn, C., Gould, A. (1992) ApJ., **389**, 234

[11] Bahcall, J. N., Schmidt, M., Soneira, R.M. (1982), ApJ. Lett., **258**, L23

[12] Bahcall, J. N., Flynn, C., Gould, A., Kirhakos, S. (1994), ApJ. Lett., **435**, L51

[13] Bennett, D. P. *et al.* (1994) in *Dark Matter*, Procs. 5th Ann. Maryland Conference, Ed. by S. Holt, in press, astro-ph/9411114

[14] Binney, J. J. (1994) in *IAU Symp. 169, Unsolved Problems of the Milky Way.* ed. by Blitz, L, Kluwer, in press.

[15] Casatano, S., van Gorkom, J. H. (1991) AJ., 101, 1031

[16] D'Antona, F. (1987) ApJ., **320**, 653

[17] Evans, N.W. (1994) MNRAS., **267**, 333

[18] Fich, M., Tramaine, S. (1991) ARA&A., **29**, 409

[19] Frieman, J., Scoccimarro, R. (1994) ApJ. Lett., **431**, L23

[20] Gould, A. (1994a) astro-ph/9408060

[21] Gould, A. (1994b) astro-ph/9409065

[22] Gould, A. (1994c) astro-ph/9403022

[23] Gould, A. (1994d) astro-ph/9412077

[24] Griest, K. (1991) ApJ., **366**, 412

[25] Griest, K. *et al.* (1991) ApJ. Lett., **372**, L79

[26] Kiraga, M., Paczyński, B. (1991) ApJ. Lett., **430**, L101

[27] Kroupa, P., Röser, S., Bastian, U. (1994) MNRAS., **266**, 412

[28] Kuijken, K., Gilmore, G. (1989) MNRAS., **239**, 605

[29] Miyamoto, S. and Yoshii, Y. (1995) in Proc. of *the Symposium on Galactic Astronomy and Space Astrometry*, p1

[30] Paczyński, B. (1986) ApJ., **304**, 1

[31] Paczyński, B. (1991) ApJ. Lett., **371**, L63

[32] Palla, F., Salpeter, E. E., Stahlar, S. W. (1983) ApJ., **271**, 632

[33] Richer, H. B., Fahlman, G. G. (1992) *Nature*, **358**, 383

[34] Rix, H.-W. (1994) in *IAU Symp. 169, Unsolved Problems of the Milky Way.* ed. by Blitz, L, Kluwer in press.

[35] Sahu, K. C. (1994) *Nature*, **370**, 275

[36] Sackett, P. D., Gould, A. (1993) ApJ., **419**, 648

[37] Scalo, J. M. (1986) *Fumd. Cosm. Phis.*, **11**, 1

[38] Sutherland *et al.* (1994) Neutrino '94, astro-ph/9409034

[39] Thomas, P. (1989) MNRAS., **238**, 1319

[40] Tyson, J. A. (1988) AJ., **96**, 1

[41] Tyson, J. A., Seitzer, P. (1988) ApJ., **335**, 552

[42] Udalski, A. *et al.* (1993) *Acta Astron.*, **43**, 289

[43] Udalski, A. *et al.* (1994) *Acta Astron.*, **44**, 165